# Solitonic State in Microscopic Dynamic Failures

H.O. Ghaffari[1]*, M.Pec[1] and W.A.Griffith[2]

[1] Department of Earth, Atmospheric and Planetary Sciences, Massachusetts Institute of Technology, Cambridge, Massachusetts, USA

[2] School of Earth Sciences, Ohio State University, Columbus, Ohio, USA.

*Correspondence to: hoghaff@mit.edu

**Abstract**: Onset of permanent deformation in crystalline materials under a sharp indenter tip is accompanied by nucleation and propagation of defects. By measuring the spatio-temporal strain field near the indenter tip during indentation tests, we demonstrate that the dynamic strain history at the moment of a displacement burst carries characteristics of formation and interaction of local excitations, or solitons. We show that dynamic propagation of multiple solitons is followed by a short time interval where the propagating fronts can accelerate suddenly. As a result of such abrupt local accelerations, duration of the fast-slip phase of a failure event is shortened. Our results show that formation and annihilation of solitons mediate the microscopic fast weakening phase, during which extreme acceleration and collision of solitons lead to non-Newtonian behavior and Lorentz contraction, i.e., shortening of solitons' characteristic length. The results open new horizons for understanding dynamic material response during failure and, more generally, complexity of earthquake sources.

**Introduction**: Apparent discontinuities in load–penetration depth (*P–h*) curves are a known characteristic of indentation tests (1-4). Such discontinuities are termed displacement bursts and usually are complemented by emission of ultrasound waves as well as charged particles and electrons (5-7). Numerical studies and post-mortem observations established that such excitations are the result of nucleation of defects, which are the carriers of plasticity and are expressed as displacement bursts in *P-h* curves (2-4). A burst represents a single or multiple slip events which relax accumulated stresses under the indenter and perturb the strain field. In particular, it has been suggested and supported by recent experiments in which the general process of splitting matter – in both gas and solid states - evolves over at least two main time scales due to distinct fast and slow relaxation processes (8-11). The fast relaxation is manifested as an abrupt decrease in load in a displacement rate controlled experiment or as a displacement burst in a load controlled experiment (hereafter we call it *weakening phase*). The weakening phase is usually followed by asymptotic relaxation to a new equilibrium state (8-9, 11). However, the dynamic role of nucleated defects in shaping the relaxation path and the mechanisms by which contact-induced plasticity leads to sharp decreases in load following a displacement bursts is still poorly understood. Here, we infer the relative change of a dynamic spatio-temporal strain field in the vicinity of the indenter tip at the onset of displacement bursts from recorded ultrasound waves. Our results show the existence of local nonlinear excitations hidden in emitted ultrasound waveforms. In particular, we uncover a peculiar mechanism in which multiple solitons propagate at different velocities but converge to collide and annihilate each other during a characteristic Λ-shaped collision. Such convergence occurs on a very short time scale and leads to fast acceleration of solitons. We show that the latter solitons clearly squeeze their characteristic length, satisfying Lorentz' contraction law.

**Results-** We indented suspended thin sheets of single crystal mica and recorded the emitted ultrasound signals (high frequency analogues to seismic waves (12-13)) using an array of 8 to 16 ultrasound sensors arranged in a ring-like pattern around the indentation site (Fig.1A; Fig.S3- see Methods section). The displacement bursts – coinciding with emission of high frequency noises – in our load-controlled experiment appear as discontinuities in the displacement record (1-4). During an indentation test of the mica sheet, we were able to record tens of ultrasound excitations well known as acoustic emissions (AEs) with different excitation levels of source energy. In Fig.1B, we show waves from a single excitation parallel to the [001] direction of a muscovite mica specimen. To analyze the array of recorded AE signals, we use a thresholded measure of correlation between sensor activities (measured in *mV*) and construct a virtual "lattice" of nodes that evolve throughout the recorded time series based on the behavior of the entire system (see Methods section for details -Fig.S1). The constructed one dimensional artificial active lattice is characterized by the average of all nodes' degree (<k> - or mean coordinate number) and evolution of the lattice topology allows us to study the complexity of the excitation sources (Fig. 1C, D). The degree $k_i$ of the $i^{th}$ node at a given time represents the number of connected links to the node where "links" are established based on a similarity metric, therefore representing the intensity of spatial correlation between the node and all other nodes (13-17). Calibration of $k(t)$ with impulsive compressive and shear sources indicates that the evolution of this parameter can be closely approximated as the relative change of dynamic strain (Methods section -Fig.S3-S4).

To visualize the temporal evolution of the dynamics of sites , we map the spatial evolution of the degree $k_i$ of the $i^{th}$ node , using polar coordinates $(R_i, \theta_i)_{i=1,...,Nodes}$ where $R_i = k_i$ and $\theta_i$ indicate the position of the node around the ring (we call this structure a *k-chain*-Fig.1E &F- Supplementary movie 1,2). It is convenient for our purposes to assign a state for each node by $s_i = sign(\frac{\partial k_i}{\partial t})$ so that $s_i = \pm 1$ -see Fig.S6. Upon excitation, the emitted waves carry information regarding the local disturbances in the vicinity of indenter tip, and propagation of such disturbances is deduced from the intensity evolution of the array-recorded waves.

Evaluating the spatio-temporal pattern of $k_i(t)$ as relative change of the strain field and its spatial mean trend $<k(t)>$, we found distinct patterns of dynamic failure characterized by (Fig.1E): (I) First, $<k>$ increases over a short time period (1 – 3 µs), (II) followed by a fast-slip phase (≈ 3-15 µs). Comparison of the Fourier transform of the recorded waveforms with $<k(t)>$- Fig.1D- shows that the onset of high frequency components coincide with the fast-weakening phase in $<k(t)>$ indicating sweeping from low energy (phase I) to high energy (II) level (Fig.S5). In phase I, the system is pulled from an equilibrium state to a state where all nodes polarize in the outward direction. This occurs in the rising part of $<k(t)>$ in Fig.1C and is stable for ~200-500 ns. After this short stable phase, the trend of correlation between nodes breaks down and one or more nodes' states reverse; the reversed node forms a kink or domain wall in the strain field and represents a perturbation to the fully correlated state which interpolate between two fully correlated-states of the system-Fig.S6. The formation of kinks is visualized by the onset of "folding" of the k-chain, and we use the general term "soliton" to describe moving kinks (Fig.1F- supplementary movie 1,2). Solitons are non-periodic waves that can be well-described by step-like functions of the form $tanh(\theta-v_{soliton}t/w)$ where $v_{soliton}$ is the propagation velocity of a soliton with a width, *w,* and position, *θ (20,21)*. From a physical point of view, a kink can be viewed as a local defect, and such defects – as we will show in detail below – affect the dynamic strain history during failure. Here we define failure as the fast drop of $<k>$ as the indicator of the mean strain. The relevance of kinks for deformation can be track down to the pioneering work by Frenkel on the shear strength of crystalline solids and later the corresponding well-studied one-dimensional Frenkel-Kontorova (FK) model (21,34), describing a 1D chain of finite coupled sites subjected to a sinusoidal potential. Processes pertaining resistance against deformation in the FK model are governed by excitation of kinks (34).

To further illuminate the evolution of moving solitons in our chain, including their interactions and resultant effects on the inferred dynamic strain field, we present the evolution of density profiles based on kinetic energy of the k-chain: $E = \frac{1}{2}\sum_{i=1}^{N} m_i (\frac{du_i}{dt})^2$ where $\frac{dk_i}{dt} \equiv \frac{du_i}{dt}$ as the rate of normal displacement (dilatational components), N is the number of sites, and we assume a non-dimensional mass $m_i = 1$ (section 2 of Supplementary Information). The above kinetic energy term only considers the radial deformation of the chain (i.e., expansion or contraction) and energies corresponding to angular deformation has been omitted. We map the angular direction along a k-chain on to a vector **q** with fixed magnitude and direction $\Theta$. Therefore, a system with N elements is characterized by **q** vectors and the energy E is defined in **q**-space ($E(\mathbf{q})$). To characterize the energy distribution in **q**-space (pseudo-momentum space), we use the concept of density of states in a 1-d chain as (20,22): $D(E) \equiv \frac{1}{2\pi |\frac{dE}{d\theta}|}$ where $\theta = \frac{n\pi}{Na}$ and n=1,..., N (a is the lattice space). If $\frac{dE}{d\theta} \to 0$, $D(E)$ will include singularities known as *Van Hove singularities-VHS* (23-24). Figure 2A shows an example of time-evolution of a k-chain in $0 \leq \theta \leq \pi$ at the onset of a soliton-antisoliton pair formation. Approaching nucleation of the soliton, the kinetic energy of the chain monotonically decreases and apparently is absorbed at certain points which are the points of soliton nucleation. In fact, the kinetic energy includes additional term regarding non-radial motion (i.e., shear deformation) due to motion of kinks. Interestingly, the nucleation points correspond to points with divergences in the density of states (VHS) and splitting of VHS (Fig.2B-inset) coincides with soliton-antisoliton propagation. In Figures 3 and 4, we show the time-evolution of *D(E)* in *θ-t* parameter space where several domain walls nucleate, propagate and finally annihilate. A soliton moves with a certain velocity $v_{soliton}$ and perturbs the initial state of the system (Fig.3,-Fig.S7). In Fig.3, we show the nucleation, propagation and interaction of solitons which occurs in the transition from phase I to phase II. During this transition, we observe that two solitons merge into a single short-lived pulse (i.e., strong merger regime (25-27)) and annihilate each other. In particular, for the shown example, four main solitons dominate transition to the fast-slip phase and they prevailingly move with the velocity of 0.03C featuring slow propagating velocity .Here, C is the maximum allowed velocity and we use a normalized value of C≡1. For mica this is on the order of a few km/s corresponding to surface waves (see Methods section for measuring C). Tracking of soliton fronts unravels the attractive interactions of colliding solitons (Fig.4-Fig.S7). This attractive interaction leads to a "sudden" increase in the front velocity up to - in less than a few hundred nanoseconds implying an abrupt local acceleration up to ~$10^9$ m/s$^2$, four orders of magnitude higher than reported in regular laboratory stick-slip experiments that employ low frequency load cells and gauges (28-29). The accelerated soliton fronts result in faster annihilation and eventually yield a shorter fast-slip phase (Fig.4); hence quickening the weakening phase of micro-failure events associated with displacement bursts. The observed Λ-shape collision of solitons and their annihilation are universal features of the studied events in our experiments and modulate the relaxation of the system across the transition between phase (I) and fast-slip (II) phases. The attractive interaction of soliton fronts may occur due to the nonlocal nonlinear nature of the medium in which the response at a certain site is transferred to the surrounding regions and induces a spatially long-range response of the medium (30-31). The lifetime of fast fronts (maximum ~ 800 ns) is shorter than the lifetime of slower fronts (maximum a few microseconds). Due to merging of the solitons, the evolution path of the system after crossing the peak of <k> is reversed and the system approaches another (ground) state in which all sites point inward (Fig.3B-Fig.S8).

A striking feature of moving solitons as proposed using several analytical solutions (20, 21, 34), is that solitons with velocities approaching the maximum allowed velocity in a given structure exhibit relativistic features. In Figure 5 –also see Fig.S9-we show the trajectory of a soliton in transition from phase I to II where a sudden jump in the front velocity of up to 6 times the initial velocity is observed. After this sudden acceleration, the velocity of the soliton approaches ~0.8C and its width *w* shrinks by ~45% of the initial length. This squeezing of soliton characteristic length is consistent with Lorentz contraction and hints at the existence of relativistic features occurring during the displacement bursts.

**Discussion -** Our results provide new fundamental insights into microscopic dynamic failure under a sharp indenter tip. We presented experimental evidence of short-term strain disturbances in the form of moving solitons in the course of displacement-bursts. Tracking the motion of single solitons unraveled their abrupt acceleration that occurs during the fast weakening phase - the most critical phase of any unstable failure. Motivated by our results, we argue that existence of solitons as the shear components to an evolving k-lattice are analogous with the observed rupture fronts moving along an active frictional interface (Fig.S3f). From this point of view, the propagated soliton along a non-deformable chain represents a single moving shear front as it has been represented in recent experiments (8,35). Therefore, we speculate that multiple nucleated fronts along an interface, in contrast with a single hypocenter, can result in a peculiar abrupt local accelerated scenario with magnitudes much higher than global observed values of the slip acceleration. Existence of a short-term relativistic state in motion of solitons implies increasing the energy of moving defects which in turn modifies the energy budget relation of the system. For a soliton moving with 0.8 C, the consumed energy is 67 % higher than for slow fronts (a few percent of C) and effectively acts as a non-trivial energy sink of the system.

In summary, our work indicates that the introduced framework to analyze recorded ultrasound excitations provides a versatile platform to study displacement bursts during contact-based failures and establish novel methods for probing dynamic defects.

**References:**

1. Schuh, C. A. Nanoindentation studies of materials. *Mater. Today* 9, 32–40 (2006).

2. Li, J., Van-Vliet, K. J., Zhu, T., Yip, S. & Suresh, S. Atomistic mechanisms governing elastic limit and incipient plasticity in crystals. *Nature* 418, 307–310 (2002).

2. Tadmor, E. B., Miller, R., Phillips, R. & Ortiz, M. Nanoindentation and incipient plasticity. *J. Mater. Res.* 14, 2233–2250 (1999).

3. Gerberich, W. W., Nelson, J. C., Lilleodden, E. T., Anderson, P. & Wyrobek, J. T. Indentation induced dislocation nucleation: The initial yield point. *Acta Mater.* **44**, 3585–3598 (1996).

4. Schuh, C. A. Mason, J. K. & Lund, A. C. Quantitative insight into dislocation nucleation from high-temperature nanoindentation experiments. *Nature Mater.* 4, 617–621 (2005).

5. Dickinson, J.T., Donaldson, E.E. and Park, M.K., The emission of electrons and positive ions from fracture of materials. *Journal of Materials Science*, 16(10), pp.2897-2908 (1981).

6. Theofanis, P.L., Jaramillo-Botero, A., Goddard III, W.A. and Xiao, H. Nonadiabatic study of dynamic electronic effects during brittle fracture of silicon. *Physical review letters*, 108(4), p.045501(2012).

7. Enomoto, Y. and Hashimoto, H. Emission of charged particles from indentation fracture of rocks. *Nature*, 346(6285), pp.641-643 (1990).

8. Ben-David, O., Rubinstein, S. & Fineberg, J. Slip-Stick: The evolution of frictional strength. Nature 463, 76 (2010).

9. Ghaffari, H. O., & Young, R. P. Acoustic-friction networks and the evolution of precursor rupture fronts in laboratory earthquakes. Scientific Reports, 3. (2013).

10. Gring, M. et al. Relaxation and prethermalization in an isolated quantum system. Science 337, 1318–1322 (2012).

11. Langen, T., Gasenzer, T. & Schmiedmayer, J. Prethermalization and universal dynamics in near-integrable quantum systems. *J. Stat. Mech.* 064009 (2016).

12. Benson, P. M., Vinciguerra, S., Meredith, P. G. & Young, R. P. Laboratory simulation of volcano seismicity. Science 322(5899), 249–252 (2008).

13. Ghaffari, H. O., W. A. Griffth, P. M. Benson, K. Xia, and R. P. Young. Observation of the Kibble–Zurek mechanism in microscopic acoustic crackling noises. *Scientific Reports* 6 (2016).

14. Newman, M. E. J. Networks: *An Introduction* (Oxford University Press, 2010).

15. Kramer, M. A., Eden, U. T., Cash, S. S. & Kolaczyk, E. D. Network inference with confidence from multivariate time series. *Physical Review E* 79(6), 061916 (2009).

16. Gao, Z., Small, M., Kurths, J. Complex network analysis of time series, *Europhysics Letters*, Vol 116, 5 (2016).

17. Bullmore, E. & Sporns, O. Complex brain networks: graph theoretical analysis of structural and functional systems. *Nature Reviews Neuroscience* **10**(3), 186–198 (2009).

18. Marwan, N., Donges, J. F., Zou, Y., Donner, R. V. & Kurths, J. Complex network approach for recurrence analysis of time series. *Phys. Lett. A* 373, 4246–4254 (2009).

19. Sethna, J. P. *Statistical Mechanics: Entropy, Order Parameters and Complexity* (Oxford Univ. Press, 2006).

20. Heeger, A. J., Kivelson, S., Schrieffer, J. R. & Su, W. P. Solitons in conducting polymers. Rev. Mod. Phys. 60, 781–850 (1988)

21. Kivshar Y. S. & Malomed B. A. Dynamics of solitons in nearly integrable systems. Rev. Mod. Phys. 61, 763–915 (1989).

22. Ashcroft, N. W. & Mermin, D. N. *Solid State Physics* (Brooks Cole, Belmont, USA, 1976).

23. Van Hove, L. The occurrence of singularities in the elastic frequency distribution of a crystal. Phys. Rev. 89, 1189–1193 (1953)

24. Kim, P., Odom, T. W., Huang, J.-L. & Lieber, C. M. Electronic density of states of atomically resolved single-walled carbon nanotubes: Van Hove singularities and end states. *Phys. Rev. Lett.* **82**, 1225–1228 (1999)

25. Khaykovich, L. & Malomed, B. A. Deviation from one dimensionality in stationary properties and collisional dynamics of matter-wave solitons. Phys. Rev. A 74, 023607 (2006).

26. Hadouaj, H., Malomed, B.A. and Maugin, G.A., 1991. Soliton-soliton collisions in a generalized Zakharov system. Physical Review A, 44(6), p.3932.

27. G. I. Stegeman & M. Segev. Optical spatial solitons and their interactions: university and diversity. Science 286, 1518 (1999);

28. Ohnaka, M. and Yamashita, T. A cohesive zone model for dynamic shear faulting based on experimentally inferred constitutive relation and strong motion source parameters. *J. Geophys. Res*. 94, 4089 (1989).

29. Chang, J. C., Lockner, D. A. & Reches, Z. Rapid acceleration leads to rapid weakening in earthquake-like laboratory experiments. *Science* 338, 101–105 (2012).

30. Snyder, A. W. & Mitchell, D. J. Accessible solitons. Science 276, 1538–1541 (1997).

31. C. Conti, M. Peccianti & G. Assanto. Route to nonlocality and observation of accessible solitons. *Phys. Rev. Lett.* **91**, 073901 (2003).

32. Bishop, A. R., and Lomdahl. P. S. Nonlinear dynamics in driven, damped sine-Gordon systems. Physica D: Nonlinear Phenomena. 18, 54-66 (1986).

33. Peyrard, M., Pnevmatikos, S.& Flytzanis, N. Discreteness effects on non-topological kink soliton dynamics in nonlinear lattices. Physica D: Nonlinear Phenomena, 19(2), pp.268-281(1986).

34. Braun, O. M. & Kivshar, Y. S. *The Frenkel–Kontorova Model: Concepts, Methods, and Applications* (Springer, 2004).

35. Nielsen, S., Taddeucci, J. & Vinciguerra, S. Experimental observation of stick-slip instability fronts. Geophys. J. Int. 180, 697 (2010).

36. Ohnaka, M. & Shen, L. F. Scaling of the shear rupture process from nucleation to dynamic propagation: Implications of geometric irregularity of the rupturing surfaces. J. Geophys. Res. B 104, 817–844 (1999).

## Materials and Methods

**Experimental procedures:**

We indented Muscovite Mica specimens (from Princeton Scientific Corp). The indentation of mica sheets was performed parallel to [001] direction. We first suspended a thin sheet of mica on a ring–like structure of ultrasound sensors and carefully glued on the sensors. Then, we mechanically exfoliated the mica by peeling sheets from the top of the specimen. We repeated this process 10 to 15 times to achieve a thin layer (~5-20 µm) of the mineral sheet suspended on the sensors. The central indentation was performed using a micro-indentation instrument (DUH-211). We used different scenarios of loading with loading rates of 3mN/s. Our focus here was on the recoded excitations during loading or creep stage of the tests. In order to detect ultrasound excitations, we utilized piezoelectric sensors with frequency bandwidth of the transducers of 0.1 to 2.5 MHz. In the primary set-up (Fig.1), we used 8 sensors; to confirm the results we employed a second set-up with 16 sensors. The acoustic (ultrasound) excitation signals are first pre-amplified at 60 dB, before being received and digitized. In Figure S.2, we have shown some of the recoded waveforms. The source of excitations are two folds: dislocations and micro-cracks. To differentiate the sources and assuming that each event is corresponding to a single point-like event, we can use a source mechanism algorithm to distinguish the nature of excitations (similar to Ref.12). While this method is an approximated method however might be supported by inferring activation energy of an event using duration of events in the rising (phase I) and weakening phase (phase II). The longer events imply the potential of micro-cracks while shorter duration of these two phases are probably related to nucleation and/or motion of dislocations. Regardless of classification of events, we use the term of "crackling" to both of the categories and note that the obtained <k>-profiles do satisfy two main component of the deformations: non- deviatoric deformations are manifested as pure expansion or contraction and mixed deformations where both isotropic and deviatoric components are present. Deviatoric components are due to excitation of (shear) solitons and when $E \to 0$ the main deformation mode of the chain is governed mainly by moving kinks (Fig.S6 and Fig.S8). To have a <k>-profile without solitonic state –and then we will have pure isotropic event- one should transit to the second phase without inducing any non-linearity to <k(t)> in vicinity of the $k_{max.}$. This is equal to push the effective temperature in thermal activation model of events to zero (T→0)-see section 3 of Supplemental Information.

**Analysis of emitted ultrasound waves:**

We map the recorded acoustic emission array (array time series for each event) to a mathematical graph (*k-transform* procedure). We use a previously developed algorithm to construct the mathematical graphs from our reordered acoustic emissions with a fixed number of nodes [16-17,20]. The main steps of the algorithm are as follows (Fig.S1): (1) The waveforms recorded at each acoustic sensor are normalized to the maximum value of the amplitude at that node. (2) Each time series is divided according to maximum segmentation, such that each segment includes only one data point. The amplitude of the *j*th segment from the *i*th time series ($1 \leq i \leq N$) is denoted by $u^{i,j}(t)$ (in units of mV). $N$ is the number of nodes or acoustic sensors. We set the length of each segment as a unit with a resolution of 25ns. (3) $u^{i,j}(t)$ is compared with $u^{k,j}(t)$ to create links between the nodes. If $d(u^{i,j}(t), u^{k,j}(t)) \leq \zeta$ (where $\zeta$ is the threshold level) we set $a_{ik}(j) = 1$ otherwise $a_{ik}(j) = 0$ where $a_{ik}(j)$ is the component of the connectivity matrix and $d(\cdot) = \|u^{i,j}(t) - u^{k,j}(t)\|$ is the employed *similarity metric*. With this metric, we simply compare the normalized amplitude of sensors in the given time-step. The employed norm in our algorithm is the absolute norm. (4) Threshold level ($\zeta$): To select a threshold level, we use a method introduced in (20, 31) and references therein) that uses an adaptive threshold criterion to select $\zeta$. The result of this algorithm is an adjacency matrix with components given by $a(x_i(t), x_k(t)) = \Theta(\zeta - |u^{i,j}(t) - u^{k,j}(t)|)$. Here $\Theta(.)$ is the Heaviside function. The constructed lattice is characterized by the average of all nodes' degree (<k>), where the degree, $k_i$, is the number of links formed between each node $i$ and all other nodes in the system. For each node we assign $s_i = sign(\frac{\partial k_i}{\partial t})$ and then $s_i = \pm 1$. With this mapping, each node in a given time step acquires one of the states ( $\rightarrow$ or $\leftarrow$ ). The employed method can be understood from the perspective of Fourier transform of equal-time correlation measure over the waveforms in a given time-window. Note that the measure of correlation functions is the root of many scattering experiments and Fourier transform of equal-time correlation function of a certain quantity –such as electron density or amplitude of observer -is the intensity of the scattered (distributed) injected energy (19). The employed algorithm quantifies the intensity of equal-time correlation between nodes; higher value of <k> indicates higher correlation in the given time; the method is more robust for a limited number of observers than the Fourier transform which is more suitable for a large number of points (18). Consequently, study of the time-evolution of k-transform of equal-time similarity patterns will show the evolution of energy distribution following a burst event. In short, in a given time interval, k-parameter represents the abundance of a certain energy level as encoded in waveforms which have been recorded by a spatial distribution of observers (nodes).

In all the reported cases in this study, we used a fixed number of sites (N=300 nodes) with interpolating the connectivity matrix, resulting the lattice constant as $a=0.2386mm$. To confirm the results, we also used a second set-up with 16 transducers where we indented a similar Muscovite Mica on a substrate of aluminum plate where the lattice parameter for N=300 nodes was $a=1.03mm$. To measure the velocities of a moving kink –which are shear solitons for the k-lattice- we track soliton movement in $\theta - D(E)$ space where we shift D(E) in a fixed time-step. We labeled the vertically shifted D(E) as "intensity" in Fig 3-5. The measured velocities are in units of $\frac{\Delta \theta}{\Delta t}$ and we convert $\Delta \theta$ (in radian) to $\Delta l$ (travel length in millimeter) based on the employed lattice spacing. The estimated velocity from this method indicate that the propagation velocity in $\theta - D(E)$ space can be higher than p-wave velocity of the mica (see Fig.S.10). This is due to the projection of velocity vectors to $\theta - D(E)$ where the mass-center of a soliton usually is in oblique angle with the circumference of the k-chain –as the θ direction in our system. Therefore should not be interpreted as hypersonic regime of propagation. We report relative velocity of a moving kink relative to maximum allowed velocity (c). To estimate the maximum allowed propagation velocity of a kink, we use the Lorentz contraction relation $\frac{\xi}{\xi_0} = \sqrt{1 - (\frac{v}{c})^2}$ where we could precisely measure the characteristic length of the soliton (ξ) before and after dramatic change of the velocity.

**Calibration of k-chains with mechanical impulsive sources:**

To associate <k(t)> with a physical source parameter, we used recorded un-amplified AE signals from known sources of impulsive compressive and shear loads recorded on a ring-like array of ultrasound transducers (Fig.S3-4). To generate compressive stress, we used a spilt Hopkinson pressure bar apparatus where an impulsive stress pulse is generated by a cylindrical steel projectile (the striker bar). The flying striker bar impacts an incident bar of identical material and diameter (Fig.S4a) and induces a strain rate on the order of ~$10^1 s^{-1}$. The source signal is transferred through the incident bar and impacts a second bar which we mounted with 6-array Piezo-electric transducers. Using Dynamic linear strain gauges and knowing that the apparatus generates compressive stresses, we can compare <k(t)> with the strain gauge records (i.e., strain and stress on the bar). Here the source has the shape of a Gaussian function with superimposed oscillations due to the resonance of the bar (Fig.S4 c). The calculated <k(t)> -phase I of the evolution-is compared with recorded strain (Fig.s4d), indicating that <k>-profiles capture main features of the stress change. To compare the <k(t)> with shear sources, we used similar unamplified ultrasound transducers as well as dynamic strain gauges and an accelerometer while two halves of the saw-cut Westerly granite slide on each other (Fig.S3). We confirmed that the calculated $<k(t)>$ represent dynamic stress change on the interface due to abrupt stress drop (stick-slip tests). Neither measured acceleration nor measured velocity represent the $<k(t)>$. The form of the $<k(T)>$ profiles measured resemble slip within granite blocks during propagation of rapid macro-rupture fronts.

**Acknowledgments.** We acknowledge discussions with B.Evan (MIT), T.Cohen (MIT) and R.Paul Young (U of Toronto).

**Author Contributions** . All authors contributed to the analysis the results and reviewed the manuscript. H.O.G and W.A.G co-wrote the manuscript. H.O.G and W.A.G designed the main tests and performed the calculations. W.A.G supervised the research and helped to analyze the results.

**Competing Interests.** The authors declare no competing interests.

**Data availability.** The data presented in the figures and that support the other findings of this study are available from the corresponding author on reasonable request. The k-chain implementation using recorded ultrasound excitations is available upon reasonable request.

**Supplementary information** accompanies this paper.

Figures:

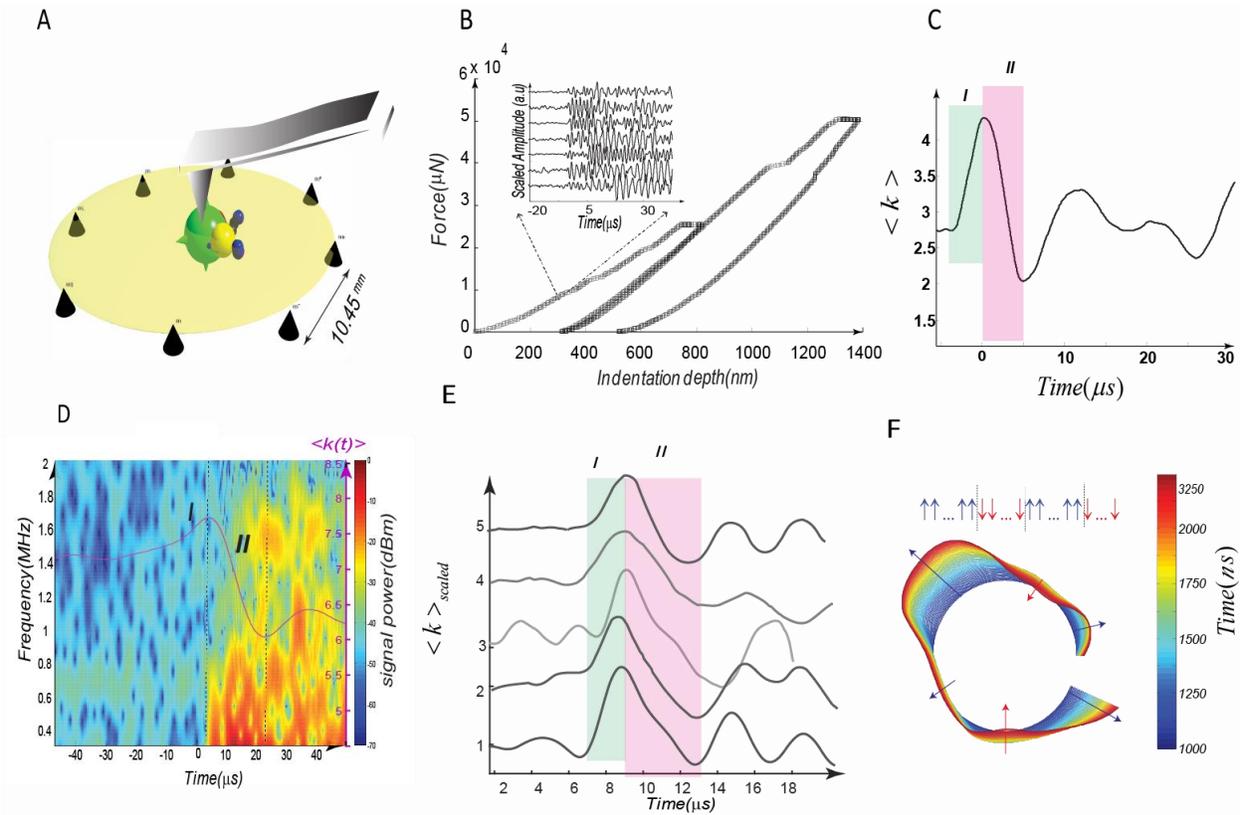

**Figure 1 Displacement bursts and ultrasound excitations in indentation of suspended Mica-films**
**(A)** Schematic representation of indentation. Thin Mica sheet is suspended over 8 piezoelectric transducers. The spheres correspond to locations where acoustic phonons are emitted as determined by a source location algorithm. The mica film covers all sensors and slightly exceeds the source-receiver distance. **(B)** Two loading-unloading paths with clear bursts events accompanied by emission of ultrasound waves. Inset shows an array of recorded excitations during the loading stage. **(C)** Evolution of the average of all nodes' degree <k> for waveforms shown in (B) showing two main stages of relaxation. **(D)** High frequency components of an excited signal in our indentation test are mapped onto fast-weakening phase in<k(t)> onset of the fast-slip coincides with the broadening of the power spectrum. An overlapped (80%) 2,048-point fast Fourier transform is used to calculate the power spectral density. In **(E)** we show 5 different acoustic events transformed to <k(t)>, where <k(t)> represents time-evolution of the mean dynamic strain field over the spatially distributed sites. After initial rising phase (Phase I), a fast relaxation phase as the sharp drop of strain (phase II) is followed. Further relaxation occurs on longer timescales. See Fig.S2 for full waveforms of the shown k-profiles. **(F)** Accumulated k-chain patterns in ~1250 ns time-interval in transition from phase I to phase II for the plot shown in (C). Also see supplementary movie 1 as a 3d visualization of accumulated k-chains.

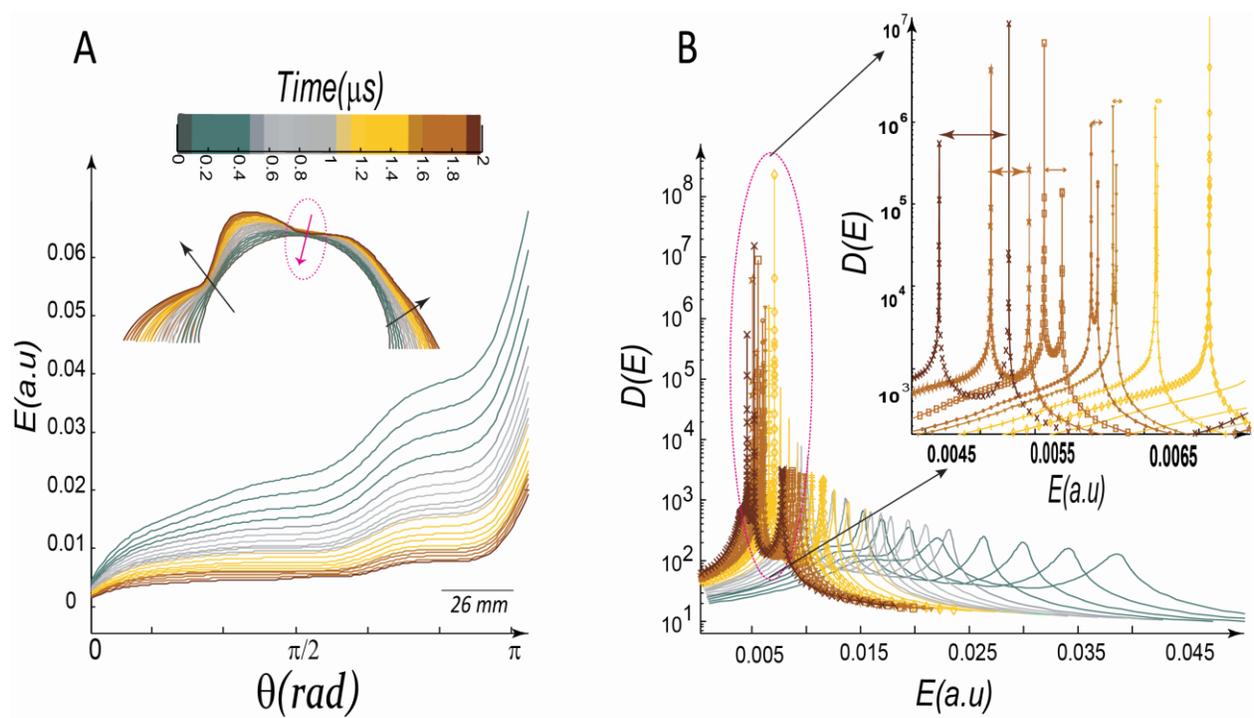

**Figure. 2. | Formation and propagation of soliton-pairs.** (**A**) The energy spectrum, E (in arbitrary units) as a function of direction Θ in different time steps. The kinetic energy of the chain decreases upon approaching nucleation of soliton shown by red arrow at the onset of folding of the k-chain. (**B**) Singularity in density profiles coincides with nucleation of a soliton pair. Inset shows the splitting between the Van Hove singularities, which is manifested in propagation of soliton-pairs. The arrows show the separation of the solitons. Approaching E→0 in (B) indicates that the moving kinks are the dominant state of the deformation of the chain.

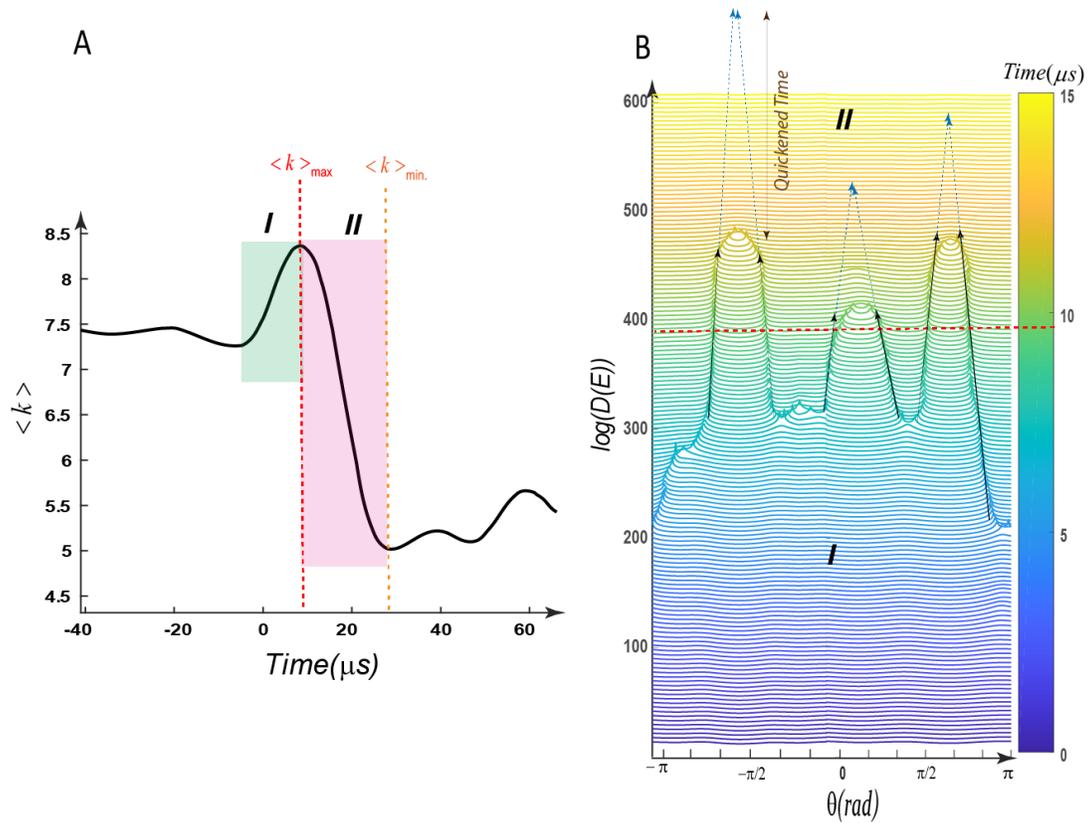

**Figure. 3.| Accelerated moving defects quicken the fast-slip phase of dynamic failures.** (A) An example of time-evolution of a k-chain as encoded in <k(t)> proportional with mean strain field. (B) Density profiles obtained between 0μs to 15 μs in 0.1 μs increments are shifted vertically for clarity. We show the hypothetical trajectories of moving kinks (dotted arrows) without acceleration which accumulatively could increase the observed duration of the weakening phase by ~12μs compared to the duration of the weakening phase with the accelerated fronts. Also see supplementary movie 2.

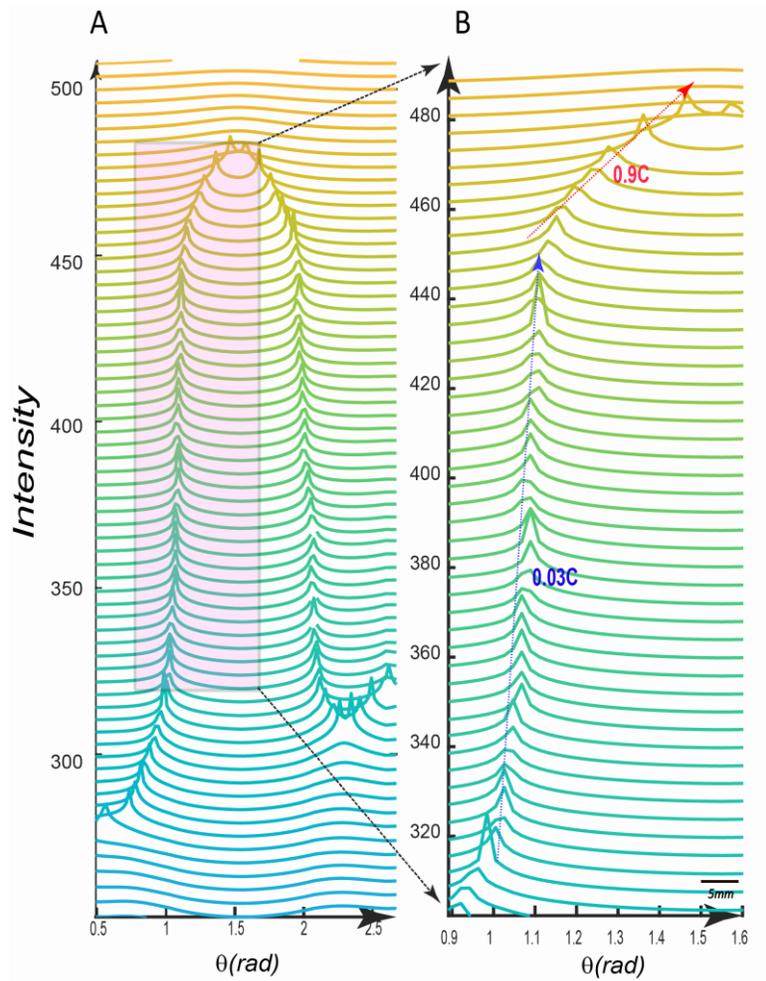

**Figure. 4.|** **(A, B) ʌ-shaped collision of two solitons**. In this example, the velocity of a soliton accelerates from 0.03C to 0.9C m/s. The ʌ-shaped collision is a universal feature of phase I→II transition. A close inspection of collisions indicates a parabolic behavior of fronts, i.e., a deviation from the linear trend. The color-code is similar to Fig.3B.

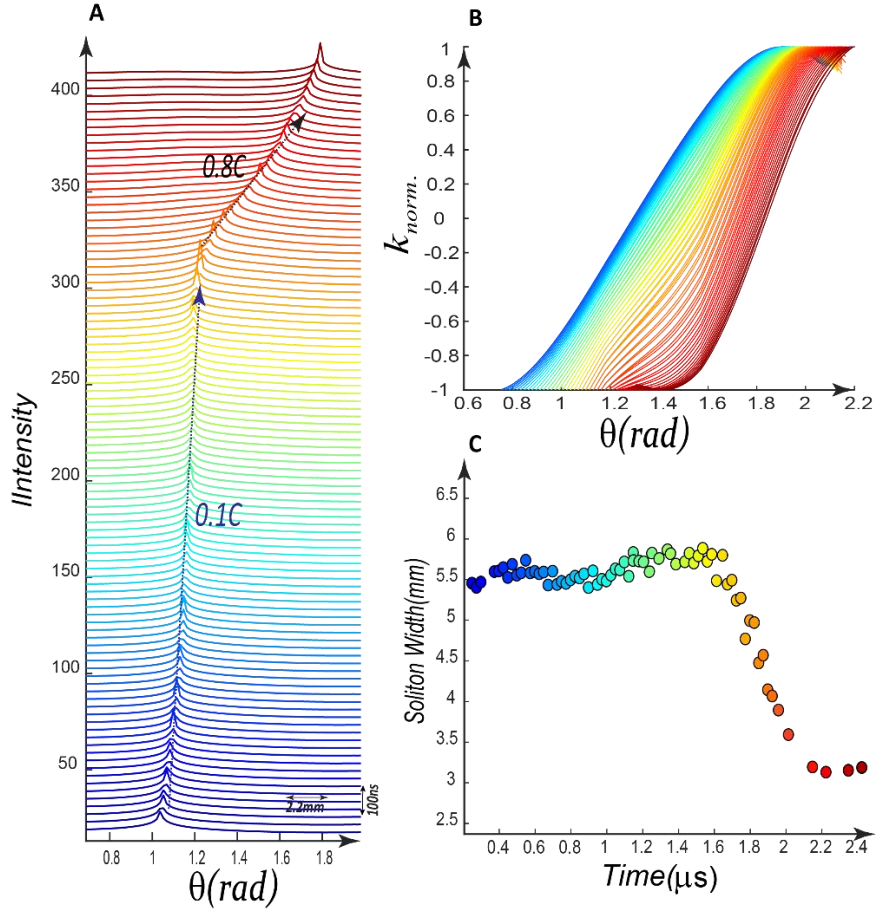

**Fig. 5. Squeezing a moving kink's width as it accelerates** **(A)** Velocity transition of a moving kink with an average velocity of 0.1C to ~0.8C in transition from phase I→II . The color denotes the passage of time. **(B)** Normalized soliton profiles with similar color coding as (a). The Lorentz contraction is evident in the profiles as a decrease in the width of the profiles (dashed arrows) when the speed of propagation $v$ is increased. We have shifted the profiles horizontally to show the effect of squeezing on the width of soliton. **(C)** The width of the soliton shrinks up to ~45% when the relative propagation velocity increases ~6 times of its initial velocity (see Fig.S9 as another example).

# Solitonic State in Microscopic Dynamic Failures

*H.O. Ghaffari [1] \*, M.Pec[1] and W.A.Griffith[2]*

[1] *Department of Earth, Atmospheric and Planetary Sciences, Massachusetts Institute of Technology, Cambridge, Massachusetts, USA*

[2] *School of Earth Sciences, Ohio State University, Columbus, Ohio, USA.*

*\*Correspondence to:  [hoghaff@mit.edu](mailto:hoghaff@mit.edu)*

# 1. Experiments

We have done several experiments using commercial ultrasound transducers (pico-sensors, and physical acoustic system-PAC-sensors). In the following, we report three different set-ups of mechanical excitation of sources. While in all indentation tests we employed amplified PZTs , in experiments pertaining impulsive mechanical loadings (compressive or shear) ,we used non-amplified PZTs. This allows us to study the recorded signals without concern on losing the data due to "clipping" of waves which usually occurs for amplified waves emitted from high energy sources.

The summary of experiments are as follows: (1) Indentation of suspended thin Mica films over 8 and 16 PZTs (Fig.S1 ; Fig.S.2 ) (2) Split-Hopkinson bar in studying compressive waves (without sample) and using dynamic strain gauges and recording axial strain field (Fig.S.4) (3) dynamic shear tests on PMMA blocks under uniaxial test and employing an accelerometer to record acceleration of the moving interface with recording un-amplified waveforms (Fig.S.3).

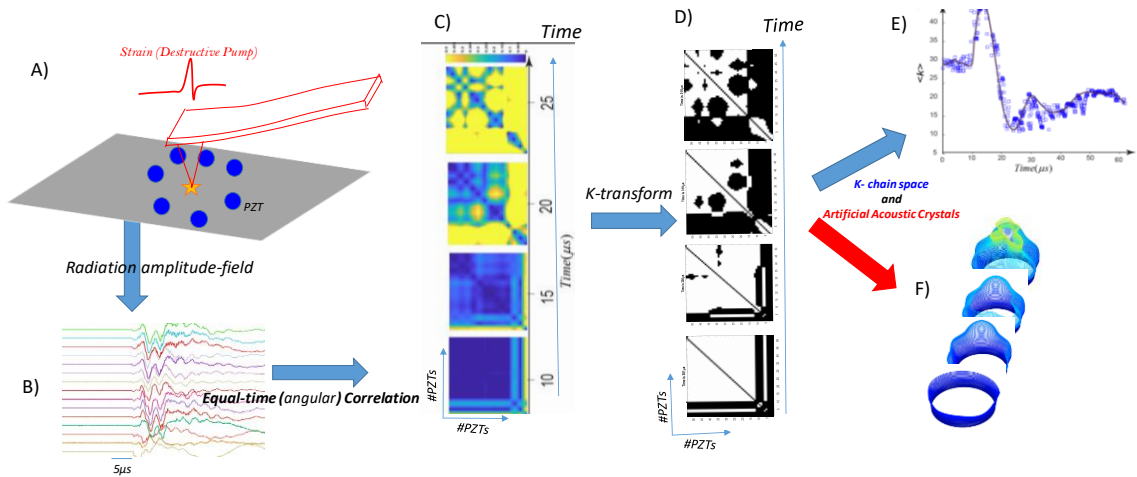

**Fig.S1. Summary of the employed method to construct k-chains from recorded ultrasound excitations. (A)** A sharp tip indents a thin-film of a mineral. The mineral could be suspended freely on transducers or on a substrate. **(B)** The recorded waveforms due to the excitation of a single event in time-window of 50 to 400 µs. **(C)** Similarity matrix in 4 successive snapshots .The most similar site-pairs are colored in yellow. **(D)** The thresholded similarity matrix as the irregular lattice configuration in the shown snapshots. **(E)** <k>-profile as an approximation to relaxation path of the system. We approximate the <k(t)> with evolution of mean strain field (see S.3 and S.4). **(F)** The constructed lattices are visualized in a ring-form with the radius at each point proportional with the number of links. With this procedure, we map the ring-structure of the PZTs into a quasi-1D (active) lattice where we could derive kinetic energy of the lattice as well as momentum-energy space of the chain.

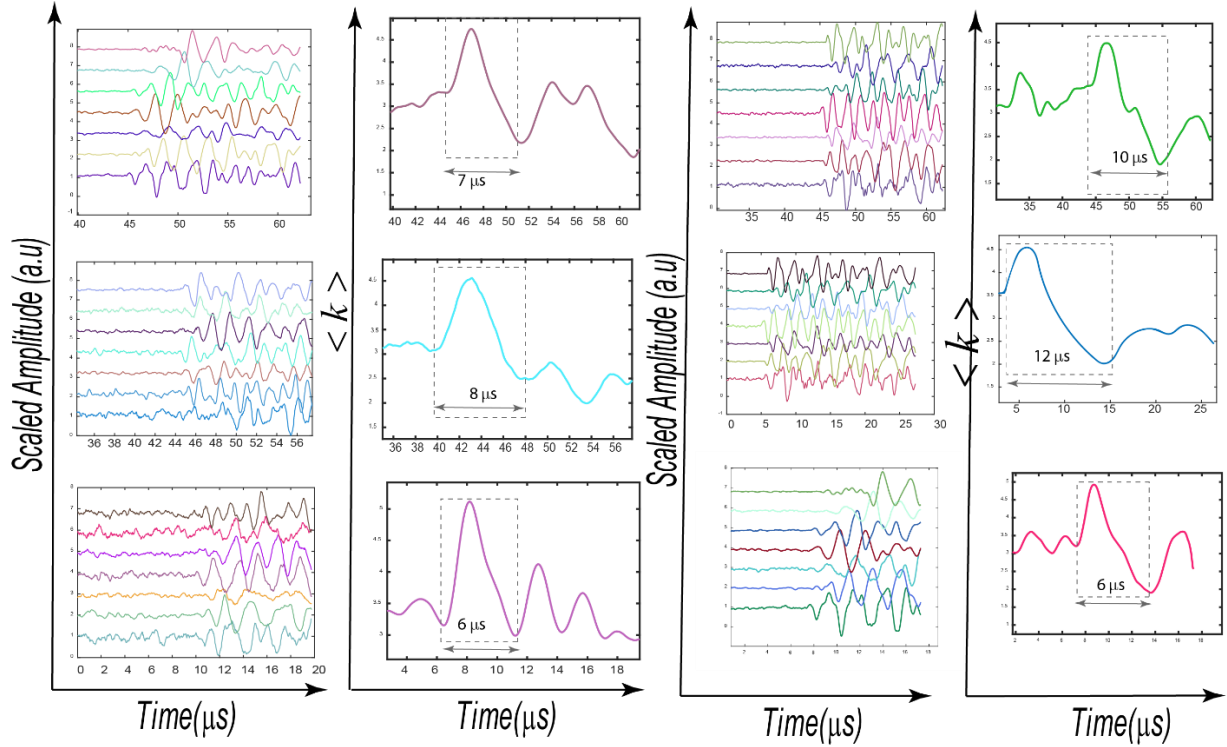

**Fig.S2.** Some of recorded waveforms of excitations in one of our experiments. We show the scaled waveforms and corresponding <k(t)> profiles . In Fig.1E of the main text, we have shown some of the <k>-parameters.

**Series of frictional experiments (shear sources):** Previously, we stablished that <k(t)> could share significant similarity to evolution of the physical recorded strain fields in stick-slip experiments in rock samples (FigS3 a-b)-Also see References [1] and [2].   An additional new experiment was carried out by employing two PMMA blocks on top of each others with an angle of ~20 degree where we used a uniaxial loading configuration on saw-cut cylinders of PMMA. We employed an accelerometer and 7 PZTs where we did not amplify the signals (Fig.S3C-F). With using simplex algorithm in determining the source location of the excited event (usually we get a single event in this case) and assuming a velocity model, we could define the source location of the triggered event. After employing k-transform procedure, the recorded waveforms were transferred in a single <k(t)> profile. Comparing the acceleration and the calculated displacement from the recorded acceleration indicated that <k(t>) does not resemble either of these parameters but is quite similar to our previous measurement on shear strain evolution on rock-rock frictional interfaces (Fig.S3a) . The Duration of the fast-weakening phase in the latter case is up to 60-70 μs while this interval is much shorter for rock-rock typical signals ~15-25 μs (Fig.S3b). Please note that the duration of fast-slip phase in most of the recoded events in our indentation tests is about ~3-12μs (Fig.S.2) much shorter than the rock-rock or PMMA-PMMA frictional tests.

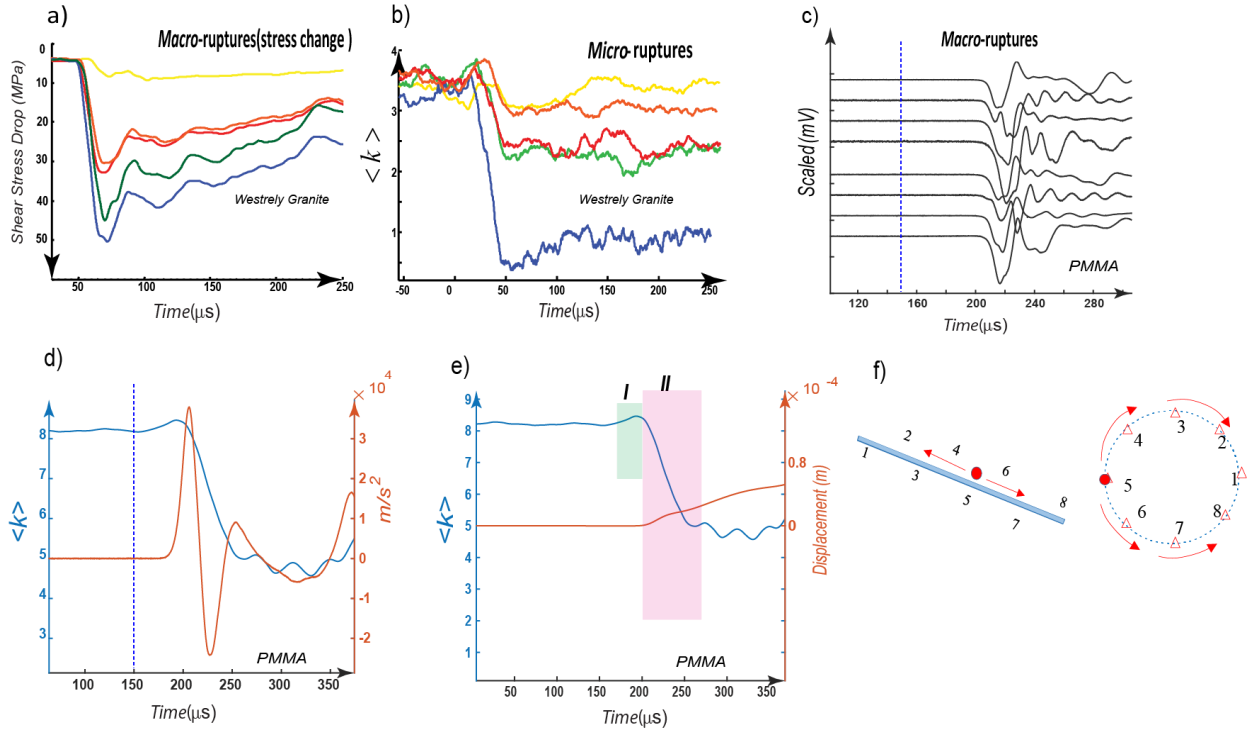

**Fig.S3. Shear source and k-chains (a)** the **measured dynamic shear stress** evolutions in two different shear-experiments : in the first case we recorded the (dynamic) strains while two halves of the saw-cut Westerly granite slide on each other leading to macro-slips (major stick-slip ). We show 5 stick-slip events with the calculated shear stresses based on the measured strains. Wheatstone bridge strain gages are employed to record the dynamic (shear) stress change at 10MHz (the slope of the fault was 60 degree with horizon (see details in [1]) **(b)** The calculated <k(t)> based on micro-cracks prior to major stick-slip experiments based on recorded ultrasound excitations [2] . The main phases of <k(t)> share similar evolutionary trends with (a) . **(c) The second shear-test: The recoded unamplified waveforms -**after stacking- under uniaxial stress while two halves of the PMMA-PMMA slide on each other. The blue dotted-line is the arrival of p-wave. **(d-e)** The calculated <k(t)> based on unamplified signals and super-imposed acceleration and displacements. The calculated <k(t)> represents dynamic strain change due to stick-slip and neither of measured acceleration or velocity represent the <k(t)>. **(f)** Mapping a frictional interface with a bi-directional rupture front on to a ring of sensors.

We can map the uni- or bidirectional rupture fronts –propagating along the frictional interface- on to a ring as we have shown in Fig.S3f. Here we virtually glue site 1 to 8 (periodic boundary condition) and then a rupture front is mapped as a rotational front along the circumference of interface-ring. We can – also-map back the orbiting fronts along the ring on an interface (Fig.S3f). Therefore, the propagated fronts along the chain on average represent evolution of an interface. Please note that here the ring is a solid interface and does not deform; i.e., it does not involve any non-deviatoric components. In the case where the ring does include isotropic terms (compressional or tensile), the deformation of the chain involves three components which could occur simultaneously.

**Compressive-impulsive sources:** In a series of experiments, first we used recorded unamplified AE signals from known sources of impulsive compressive recorded on an array of ultrasound transducers; (Fig.S.4). Using the spilt Hopkinson pressure bar apparatus, an impulsive stress pulse is generated by a cylindrical steel projectile (the striker bar). The flying striker bar impacts an incident bar of identical material and diameter (Fig.S4.a). The source signal is transferred through the incident bar and impacts a second bar which we mounted with 6-array Piezo-electric transducers. Using dynamic linear strain gauges and knowing that the apparatus generates compressive stresses, we compare the sensors response with the strain gauge records (i.e., strain and stress on the bar). Here the source has the shape of a Gaussian function with superimposed oscillations due to the resonance of the bar (Fig.S4b). The experiment is similar to employing amplified sensors in higher energy ball-drop tests in order to calibrate PZTs where a Hertzian contact force is assumed as the source and the response of the instrument is filter out (convolving) with assuming the response of the media. As we have shown in Fig.S4b the maximum amplitude of the first motion of the PZT coincides with the maximum linear strain –prior to distortion of the signal by the resonance waves of the vibrating bar. The compressive loading portion in the impulsive source is recoded with an initial rising phase in the PZT's signal confirming that the employed PZTs are positively polarized to compressive stresses. Then, the calculated $<k(t)>$ -phase I of the evolution-is compared with recorded strain (Fig.S4d), indicating that $<k>$-profiles capture main features of the stress change.

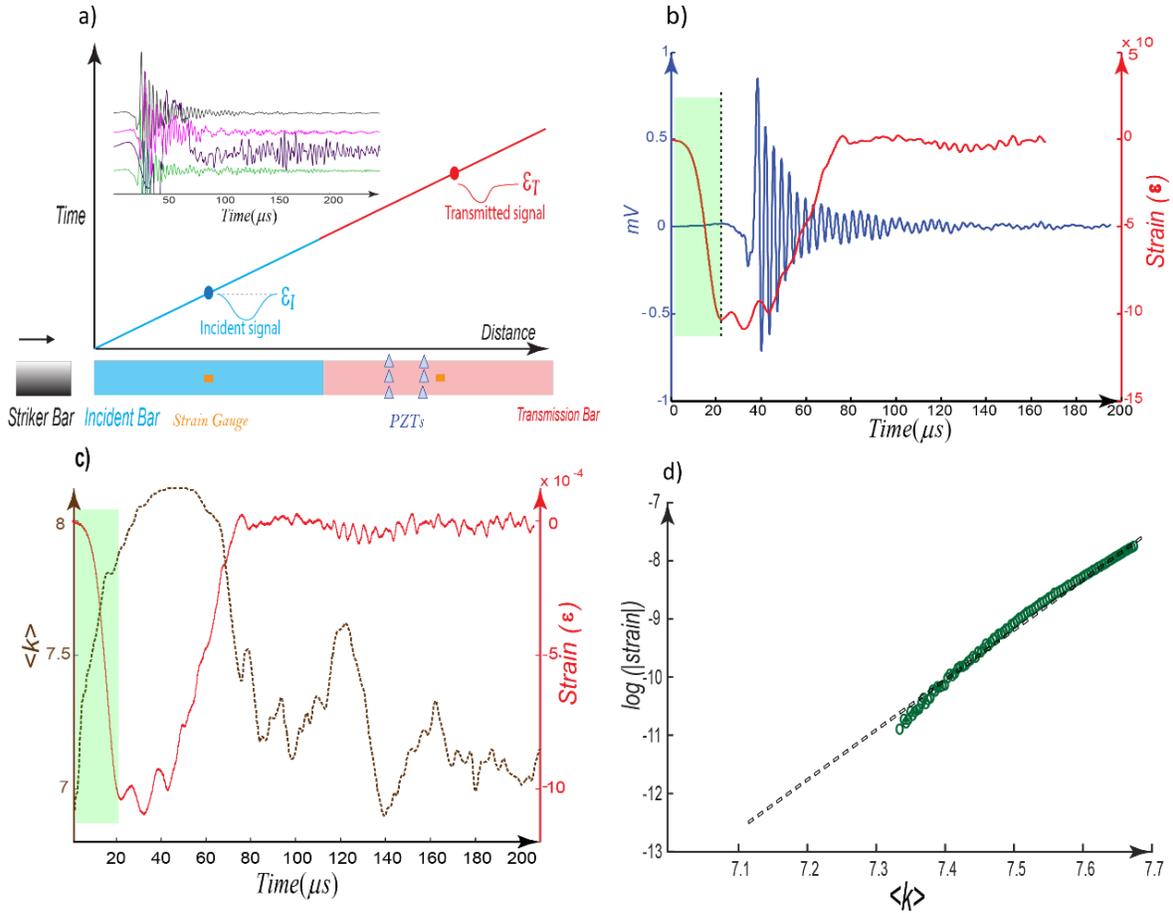

**Fig. S4. Compressive Source and k-chains.** Calibration of the employed ultrasound transducers with known source and (dynamic) strain gauges using impulsive dynamic loading tests (generated by split Hopkinson Pressure bar). (a) Here we show the response of the sensors to a controlled source tuned by the velocity of striker bar and pulse shaper. We used two dynamic strain gauges in incident and transmitted bar. We also used 6 unamplified piezo-electric transducers (PZTs) with the frequency bandwidth of the transducers from 0.2 to 1.3 MHz. The transducers were mounted on transmission bar in different positions. (b) Transmitted compressional signal (in red and negative sign of strain) superimposed on the recorded PZT signal. (c) Double plot of the recorded strain gauge and calculated <k> parameter. (d) Semi-logarithmic scale of<k> vs. strain during the rising phase of strain-time prior to effects of bar-resonance.

## 2. On Energy of k-Chains

In the main text, we evaluated the quasi momentum- energy space where we used kinetic energy term and ignored other energy terms. Here we point out some other possibilities in estimation of interaction (potential) field and using spectral analysis to analysis the introduced (active) 1-d lattices. In Fig.S.5, we show that the transition from phase I (as the initial strengthening phase) to the second phase (weakening phase) is accompanied with transition to higher energy (higher frequency component) of the waveforms. In particular, in Fig.S5b, we see clear energy bands in $<k>$- *energy* parameter space for three different events. Here we have calculated the signal power (from Fourier transformation analysis) averaged over the frequency range of 0.1MHz -2MHz . The transition zone with energy gap of $2\Delta$ is the manifestation of the solitonic state. One can assume that $<k>$ is the local-control parameter of the system (=k-chain) and is ramped from $<k>_0$ (initial value) to $<k>_I$ (see Ref.13 of the main text).

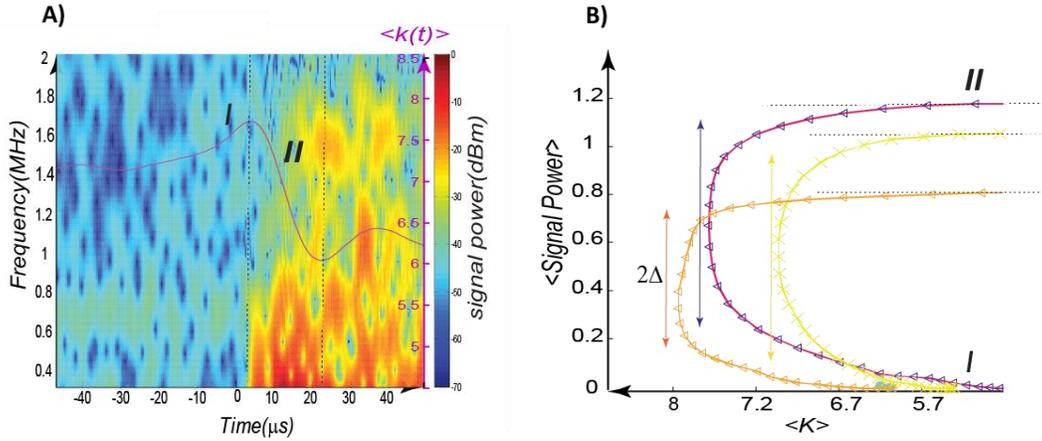

**Fig. S5. A) Study of energy through spectral analysis versus <k> in a given waveform. B)**

In general we can write the whole energy of the system (k-chain) as follows:

$$E = \frac{1}{2} M \sum_{i=1}^{n} (\frac{du}{dt})^2 + \frac{1}{2} \sum_{i=1}^{n} K(u_{n+1} - u_n)^2 + (...),$$

The first term in right hand of the equation is the kinetic energy term, the second term is interaction energy (potential) energy between the neighborhood sites and the last term is the possible interaction with other fields (such as electrons). In the main text we assumed that the kinetic energy is the dominant term of the energy and we build energy-quasi-momentum space (energy bands) of our 1D lattice based on this assumption. However, one can deduce a potential field $\Psi(r_{ij})$ for relative stable fields. Here the stable phases are pure expansion or contraction of the k-chain occurs in the first or the second phase. Assuming an isotropic source, the pattern of the k-chain at the stable phase represents the movement of each site from an initial state (the initial state is the noise level of the system). We can infer the relative potential field with assigning a stretched (or compressed) spring with certain stiffness (Fig.S6). In Fig.S.6, we have represented the potential field of the chain at the end of the evolution of the first phase where we simplified the pattern as the symmetric pattern centered around the minimum of Δk=k-k$_0$ located around θ=π. Assuming a single type of the spring (bonds) connecting the sites, we can infer a relative potential field based on the magnitude of Δk (Fig.6c); higher Δk indicates larger stretch of bonds and vice versa. At the end of the first phase, the initial equilibrated chain with $a_0$ is deformed and a new lattice-pattern forms with two (idealized) lattice parameters $a_1$, $a_2$. Here we have simplified the system with a symmetric configuration around minimum of Δk. With assuming $K \equiv 1$ (stiffness parameter) and relative displacement of the site from an initial state, we can estimate the potential field. Alternatively, one can use two types of bonds on an idealized configuration of the first or the second pure phases (Fig.S6B). Here each "unit cell" of the chain is made of two types of springs. Further extension of the energy term will be presented somewhere else.

While we reported some of the main features of the transferred waves in k-lattice space, we did not address some other features; these include transitions in velocities, weak interaction regime of solitons, bound-state of solitons, possible role of interactions of released charged particles -due to dethatching of molecular bonds- with sound waves (phonons) and its effect on velocity or path of the solitons.

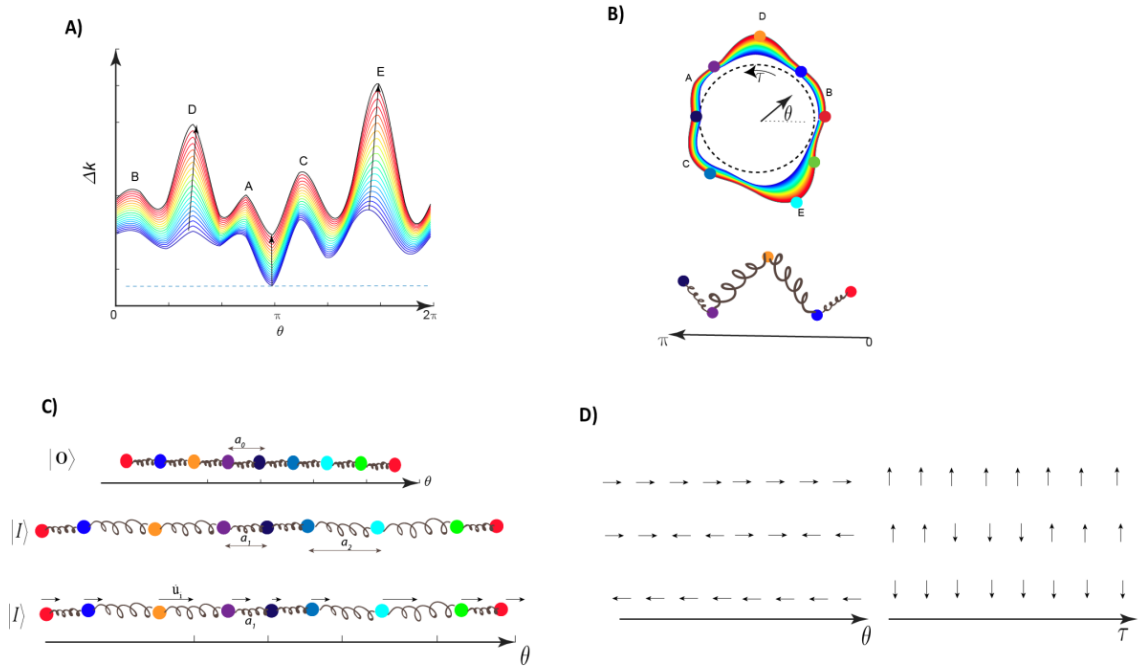

**Fig. S6.** Relative change of Δk at the end of the evolution of the first phase prior to the onset of solitonic state. The colors indicate the time and the dotted black circle is the initial state. In (B) we show the final state at the first phase. The configuration can be idealized with employing two-types of springs which has been configured in an exaggerated in zig-zag form for 0<θ<π. In (c), we show the idealized configuration of the chain using the spring-mass system. Each mass –prior to onset of the solitonic state- has the components of displacement and velocity parallel to the θ-direction. At the end of the first phase, the initial equilibrated chain with $a_0$ is deformed and a new lattice-pattern forms with two (idealized) lattice parameters $a_1$, $a_2$. Here we have simplified the system with a symmetric configuration around minimum of Δk. The arrows in (c) show the relative velocity of expansion. (D) we show two main deformation modes of the k-chains over uniform deformation : the inward or out-ward (expansion or contraction) deformation (i.e., θ coordinate) .A chain could have the mixed of both modes. The circumferential deformation (shear) (i.e., τ -coordinate) is solitonic state interpolating between two main expansion and contraction modes. We could have clock-wise and counter-clockwise mixed modes. For pure shear mode-state of a k-chain , we will not have isotropic components. One can see that the evolution of a k-chain is transition from pure θ –coordinate deformation to τ –coordinate where the energy of expansion or contraction modes will limit to zero (Fig.8b).

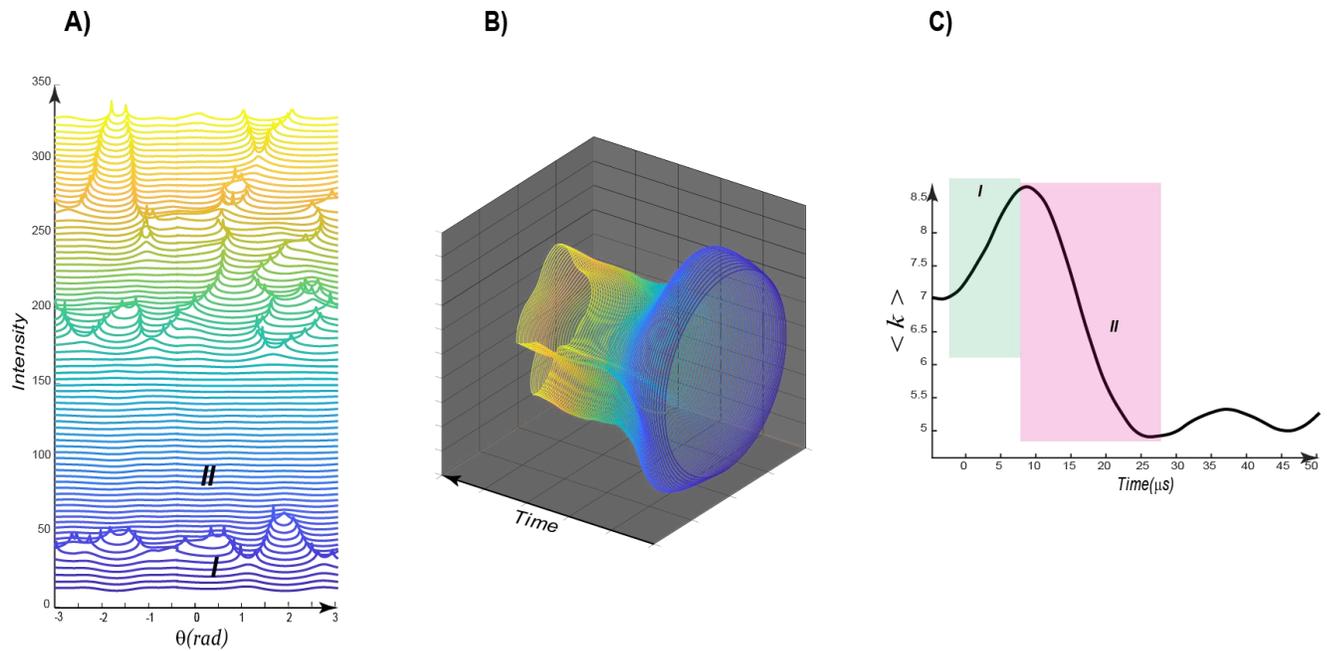

**Fig.S7. 3D visualization of a k-chain's evolution** (see the supplementary movies 1 and 2).

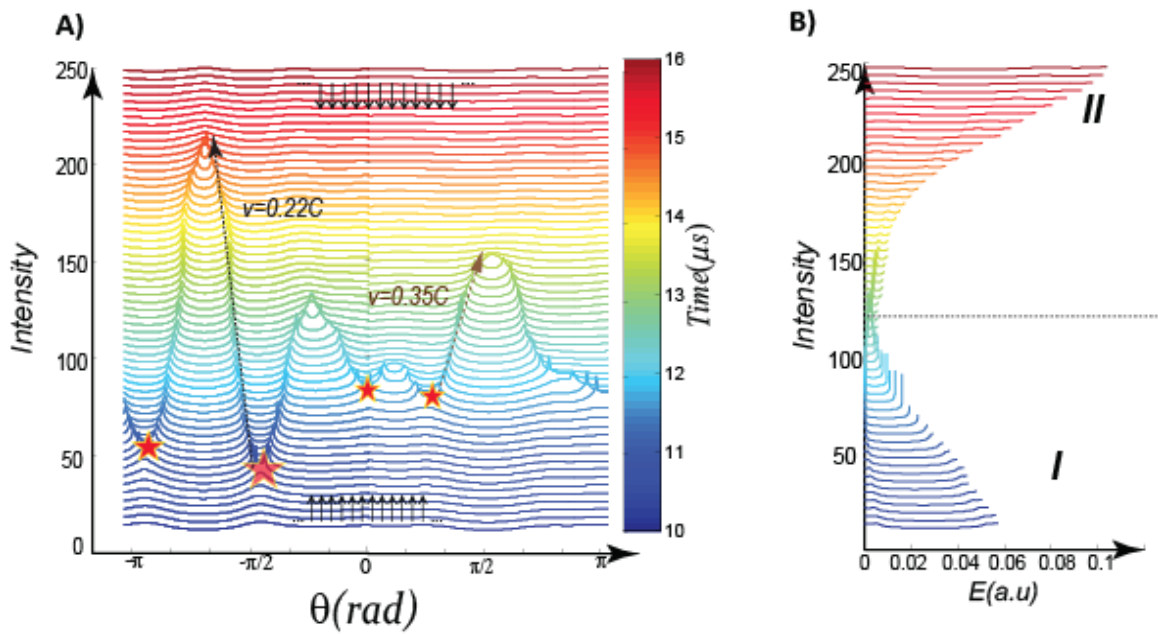

**Fig.S8. Formation of 4 moving soliton-pairs** along the chain in transition of phase I to phase II. One clearly observes proliferation, colliding and merging of fronts. E is the kinetic energy of shrinking (folding) or expanding sites. Approaching E→0 in (B) indicates that the solitons (=moving kinks) are dominant state of the deformation of the chain.

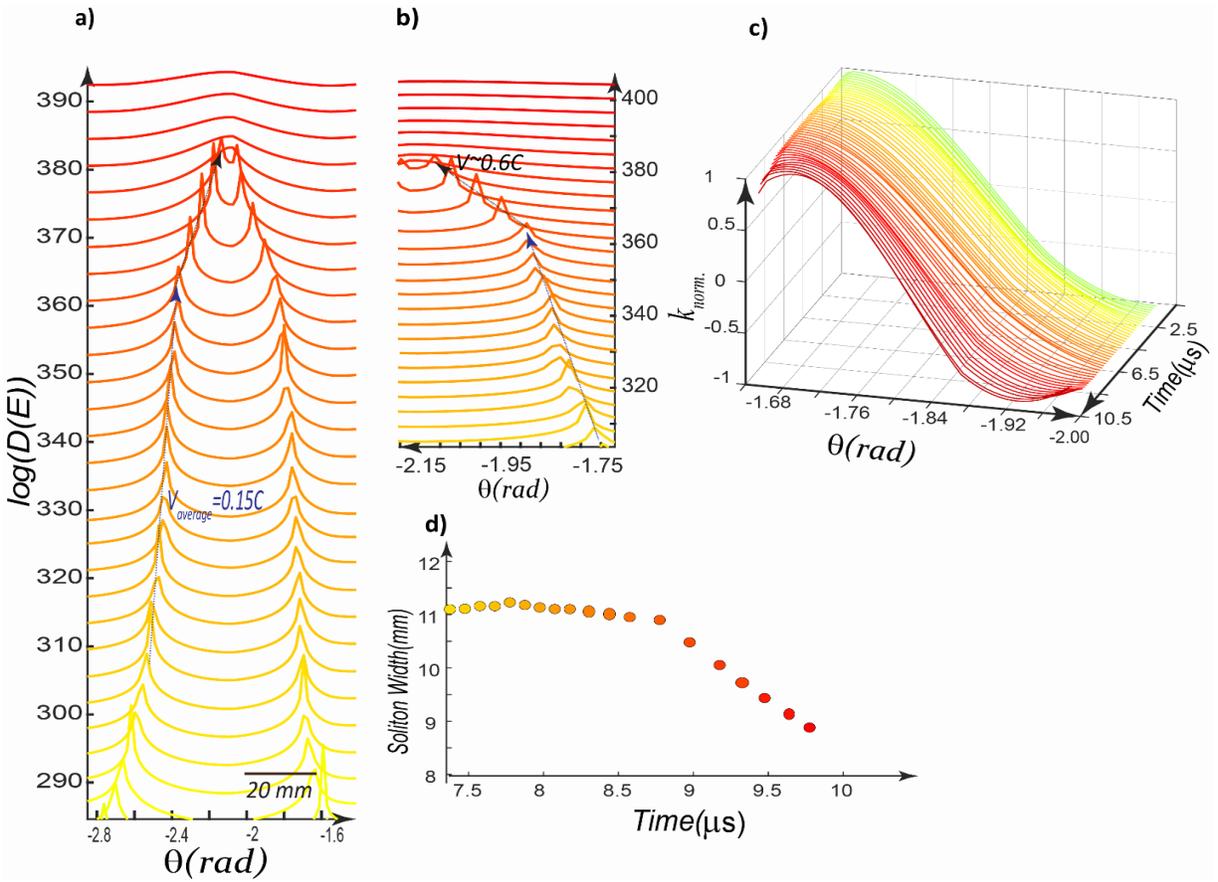

**Fig. S9| Observation of Lorentz contraction in an accelerated soliton. A,B)** Velocity transition of a moving kink with velocity of ~0.15C to ~0.6C in transition from phase I→II, leading to a fast acceleration. The color denotes the passage of time. The density profiles are in 0.1 µs increments and are shifted vertically for clarity. **C)** Normalized soliton profiles with similar color coding as (A&B). The Lorentz contraction is evident in the profiles as a decrease in the width of the profiles when the speed of propagation is increased. **D)** The width of the soliton shrinks when the relative propagation velocity increases ~4 times its initial velocity.

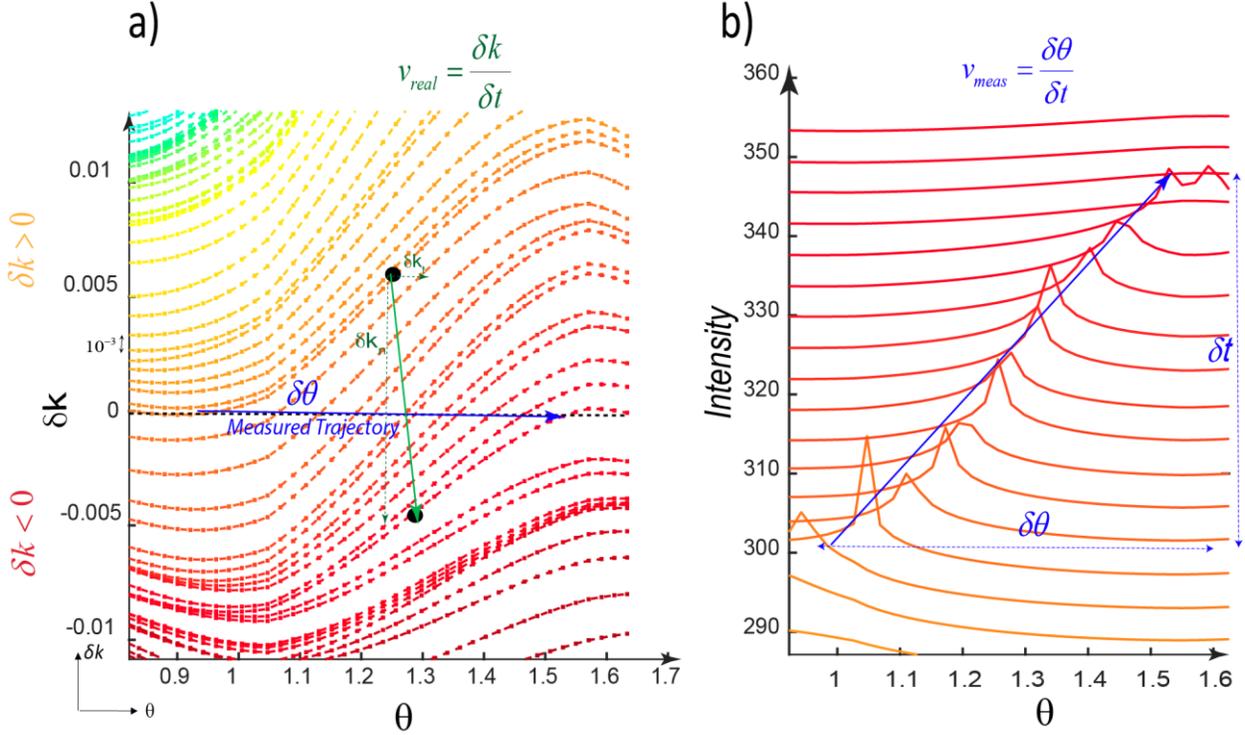

**Fig. S10| Solitons:** We define a soliton when a profile of $\delta k(\theta)$ crosses $\delta k = 0$. The rate of moving "zero" modes, $\delta k = 0$, has been reported as soliton velocity ($v_{meas} \equiv \frac{\delta \theta}{\delta t}$). The real soliton velocity is given by $v_{real} \equiv \frac{\delta k}{\delta t}$ which usually is smaller.

## 3. K-lattices : Relationship with *Fermi-Dirac* distribution and interatomic distance-force curves

Here we provide a semi-theoretical proof of k-chains as a measure of (mean) strain field. To do this, we map the k-chains onto a free electron gas model and use *Buehler-Gao's* [3] interpretation of interatomic force (hereafter B-G's model). To start we assume a configuration of a k-chain prior to onset of a kink excitation. The clusters –or communities [Ref. 14, 20 of the main text]- of the similar sub-energies carry out the energy of the chain; hence we define a certain number of the main energy levels (*m*) and as a first order approximation, we consider non-interactive energy levels. Each energy level does include sub-level energies which through the links form a community; in non-interactive energy clusters the links are purely within main energy levels and do not extend to other energy levels ("*ideal gas*" analogy). We assign a kinetic energy to each level $\varepsilon_i$ and abundance of the $i^{th}$ state, i.e., $k_i$, which is proportional with the occupation number of the $i^{th}$ level $n_i = \frac{k_i}{\sum_m k_i}$ and summation goes over *m*-main energy levels. The $k_i$ is, then, number of within links which is established between sub-levels of $i^{th}$ level. Next, We use a probabilistic argument that probability of finding an energy level with a higher value declines exponentially proportional with the energy level. Furthermore, we can define a maximum kinetic energy μ as a reference point; this is the allowed maximum kinetic energy level (in analogy with the Fermi level). The occupation number of $i^{th}$ energy level is, therefore, given by the Fermi-Dirac distribution [4]: $n_i = \dfrac{1}{e^{(\varepsilon_i - \mu)/k_B T} + 1}$ in which $n_i$ is the occupation number of the $i^{th}$ energy level, $\varepsilon_i$ is the kinetic energy, µ is the internal chemical potential (at zero temperature, this is the maximum kinetic energy, i.e. Fermi energy), T is the absolute temperature $k_B$ is the Boltzmann constant. Now, we relate the above argument to stress-strain curve (i.e., force-distance). To this end, we use *Buehler-Gao's* interpretation of interatomic force versus atomic separation $r$ which is given by (3): $F(r) = k(r - r_0)\left[\dfrac{1}{e^{\Xi(r-r_c)} + 1}\right]$ where the parameter $r_0$ refers to the nearest-neighbor spacing of atoms. Assuming that the spring constant $k$ is fixed, the $F(r)$ has two other free parameters, $r_c$ and $\Xi$. The parameter $r_c$ corresponds to the Fermi energy in the Fermi–Dirac function µ and denotes the critical separation for breaking of the atomic bonds. The parameter $\Xi$ corresponds to the temperature in the Fermi–Dirac function and describes the intensity of smoothing at the breaking point. The exact mapping can be achieved by [3] $r \leftrightarrow \varepsilon_i, \Xi \leftrightarrow \dfrac{1}{k_B T}, r_c \leftrightarrow \mu$. Thus, the evolution of $i^{th}$ energy level in a k-chain is related to interatomic separation as follows: $r_i = r_c + k_B T \log(\dfrac{\sum_m k_i}{k_i} - 1)$ which indicates that the growth of the energy levels are exponentially proportional with the interatomic separation and therefore cohesive stress (Fig.S.11). Interestingly, the later conclusion confirms our calibration results upon to the failure point in which a semi-logarithmic change of the strain is scaled with the variation of node's degree (Fig.S.4d). It is noteworthy that increasing effective temperature in B-G's model- note that thermal energy is Boltzmann's constant $k$ multiplied by temperature $T$- is in direct connection with the smoothness of the stress-strain curve in vicinity of the failure point; increasing effective temperature yields smoother curvature in failure point. Knowing that the solitionc mode -including number of solitons and their interactions -significantly shapes the smoothness of the failure curve, therefore ,we assign the effective temperature to a measure of soliton state in <k>-t curves. This inference suggests that soliton can be used equally as measure of effective temperature.

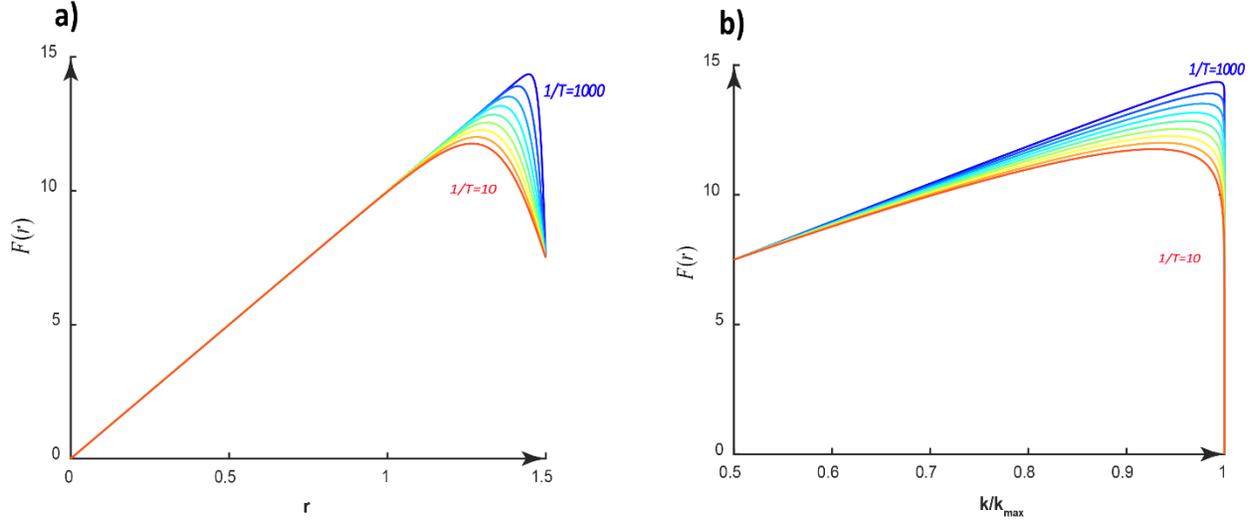

**Fig. S11|** a) cohesive force versus interatomic separation based on *Buehler-Gao's model* for different temperatures. Approaching T→0, the failure curve is more abrupt. b) Variation of interatomic separation versus a state of a site ($k_i$) and for different $k_BT$ with $k_B=1$. We have used $r_i = r_c + k_B T \log(\frac{\sum_m k_i}{k_i} - 1)$ with $r_c = 1.5$ is the maximum allowed separation of the bonds (i.e., energy barrier in terms of thermal activation) and *m* is the number of energy-levels (communities in the k-chain).

## *References*:

[1] Thompson, B. D., Young, R. P. & Lockner, D. A. Premonitory acoustic emissions and stick-slip in natural and smooth-faulted Westerly granite. *J Geophys Res*. 114, B02205J (2009).

[2] Ghaffari, H. O., Nasseri, M. H. B. & Young, R. P. Faulting of Rocks in a Three-Dimensional Stress Field by Micro-Anticracks. *Scientific Report*, 4 (2014).

[3] Buehler, M.J. and Gao, H., 2006. Dynamical fracture instabilities due to local hyperelasticity at crack tips. Nature, 439(7074), p.307.

[4] Sethna, J. P. *Statistical Mechanics: Entropy, Order Parameters and Complexity* (Oxford Univ. Press, 2006).